\documentstyle[preprint,aps,epsfig]{revtex}

\begin{document}

\begin{center}
{\Large {\bf Spin and energy transfer in nanocrystals without
transport of charge}}

\vskip 0.1cm

\ \\ \ A.~O.~Govorov \ \\ \ {\it Department of Physics and
Astronomy, Ohio University, Athens, Ohio 45701, USA}

\vskip 0.7 cm {\bf ABSTRACT} \vskip 0.3 cm

\end{center}

We describe a mechanism of spin transfer between individual
quantum dots that does not involve tunneling. Incident
circularly-polarized photons create inter-band excitons with
non-zero electron spin in the first quantum dot. When the
quantum-dot pair is properly designed, this excitation can be
transferred to the neighboring dot via the Coulomb interaction
with either {\it conservation} or {\it flipping} of the electron
spin. The second dot can radiate circularly-polarized photons at
lower energy. Selection rules for spin transfer are determined by
the resonant conditions and by the strong spin-orbit interaction
in the valence band of nanocrystals. Coulomb-induced energy and
spin transfer in pairs and chains of dots can become very
efficient under resonant conditions. The electron can preserve its
spin orientation even in randomly-oriented nanocrystals.

\newpage

Manipulation of spins in nanostructures is presently attracting a
tremendous amount of interest \cite{spins1,spins2,Ganichev}. Since
spins in solids have relatively long lifetimes, they can be
exploited as qubits - basic elements of quantum computer
\cite{QComp}. Spin-polarized states of electrons in crystals can
be generated optically
\cite{optical-orientaion,Dyakonov,Awschalom1}, by driving current
through spin-dependent barriers, or by injecting electrons from
ferromagnetic materials \cite{spins1,spins2}. In most cases, spin
transport across a crystal occurs either via tunneling or
injection. This would not be the case for so-called colloidal
quantum dots (QDs), where individual nanocrystals strongly confine
carriers and do not permit efficient tunnel coupling
\cite{exper1,exper2,exper3}. However, instead of direct tunnel
coupling, the colloidal QDs permit long-range Coulomb-induced
transfer of optically-excited excitons
\cite{exper1,exper2,exper3}. Such transport has been observed in
several recent experiments and is often referred to as F\"{o}rster
energy transfer \cite{forster}. Theoretically, F\"{o}rster-like
transfer in nanocrystals has been discussed in connection with
exciton dynamics in QD arrays and quantum computing \cite{theory}

Here we develop a theory of electron spin transfer between
individual nanocrystals without tunneling, involving
optically-excited excitons and the Coulomb interaction. So far,
spin transport in nanostructures has been considered almost
exclusively in relation to direct transport of charge
\cite{spins1,spins2}. Since the spin orientation in the conduction
band of semiconductors can be efficiently created with the
circularly-polarized light pumping 
\cite{optical-orientaion,Dyakonov}, it is interesting to study a
possibility of spin transfer between individual dots without
transfer of charge. In such a transfer process, the optical and
spin selection rules would be dictated by the strong spin-orbit
interaction in the valence band. The typical experimental scheme
related to F\"{o}rster transport involves pairs of quantum dots
with different sizes (fig.~1a-c). An incident photon creates an
exciton in the small dot 1 with a larger optical gap (fig.~1a-c).
Then, the exciton is transferred via the F\"{o}rster-like
mechanism into the large dot 2 with a smaller optical gap. Due to
fast energy relaxation in the dot 2, the exciton becomes trapped
and contribute to the photoluminescence (PL) at the dot-2 energy.
If electrons in the dot 1 are created by circularly-polarized
light, they become spin-polarized due to the spin-orbit
interaction in the valence band
\cite{optical-orientaion,Dyakonov}. Here we will focus on dynamics
of excitons generated by circularly-polarized photons and develop
principles for electron-spin transport in QD pairs without
tunneling. We will show that the spin orientation can be
efficiently transported between QDs via the Coulomb interaction.
This becomes possible thanks to the strong spin-orbit interaction
in the valence bands of QDs. The spin-transfer selection rules
strongly depend on geometry and resonance conditions. In the
resonance regime, the transfer can lead to either conservation or
flipping of spin.

Pairs of semiconductor QDs can be grown by using self-organization
technology \cite{stachedQDs}. In such stacked QDs, sizes of dots
and inter-dot separation are well controlled. Another method to
fabricate a system with QD pairs is colloidal synthesis
\cite{exper1,exper2,exper3}. In a solid of colloidal QDs with two
distinct sizes, QD pairs are randomly oriented
\cite{exper1,exper2}. In monolayers of QDs (fig.~3a), the
orientation of pairs is directional \cite{exper2,exper3}. Another
possibility to avoid a randomness is to study a single QD pair
bound to a surface \cite{exper3,Bawendi}.

In what follows, we will use several simplifications related to
the time scales. In particular, we will assume that
$\tau_{e-spin},\ \tau_{exc}\gg\tau_{energy},\ \tau_{h-spin}$,
where $\tau_{exc}$ is the exciton lifetime in a single QD related
to radiational and non-radiational transitions, $\tau_{energy}$ is
the energy relaxation time of excitons within a dot, and
$\tau_{e-spin}$ is the electron spin lifetime, and $\tau_{h-spin}$
is the momentum relaxation time of holes. In other words, we
suppose: (1) fast intra-dot relaxation of angular momentum of
holes and (2) fast energy relaxation to the ground state in the
dots.

{\it Disk-shaped dots with a cubic lattice.} First we consider a
pair of oblate (disk-shaped) quantum dots (fig.~1) with dimensions
$a_i\ll b_i$, where $a_i$ is the QD size in the $z$-direction,
$b_i$ is the in-plane diameter, and $i$ is the dot index
($i=1,2$). For simplicity, we assume that the QD potential has
infinite walls. In such a model, a single QD is
quasi-two-dimensional (2D) and its valence-band structure is
similar to that in a 2D quantum well \cite{Ivchenko}. To find the
wave functions, we first quantize the motion of heavy and light
holes in the z-direction; it provides us with the Bloch functions.
Then we can introduce weak quantization in the $x-y$ plane
involving effective masses of holes. The wave functions in the
conduction and valence bands for the dots 1 and 2 take a form:

\begin{eqnarray}
\label{WF1}
\Psi_{i}^{e,\uparrow(\downarrow),n,l}=u_{\uparrow(\downarrow)}\Phi^{(i)}_{n,l}(r_{||},z),
\ \Psi_{i}^{hh,\pm 3/2,n,l}=u_{\pm 3/2}\Phi^{(i)}_{n,l}(r_{||},z),
\ \Psi_{i}^{lh,\pm 1/2,n.l}=u_{\pm 1/2}\Phi^{(i)}_{n,l}(r_{||},z),
\end{eqnarray}
where $i=1,2$ and ${\bf r}_{||}=(x,y)$;
$u_{\uparrow(\downarrow)}$, $u_{\pm 3/2}$, and $u_{\pm 1/2}$ are
the Bloch functions of electrons, heavy holes ($hh$) and light
holes ($lh$), respectively;
$\Phi^{(i)}_{n,l}(r_{||},z)=f^{(i)}_0(z)R^{(i)}_{n,l}(r_{||})$ are
the envelope functions, where the $f^{(i)}_0(z)$ is the
ground-state function for the motion in the $z$-direction,
$R^{(i)}_{n,l}(r_{||})$ are Bessel's functions describing the
in-plane motion, and $(n,l)$ are the radial and azimuthal quantum
numbers of in-plane motion, respectively; $n=1,2,...$ and
$l=0,\mp1,\pm2,...$.  In our simplified approach all types of
carriers are described with the same set of envelope wave
functions $\Phi_{n,l}(r_{||},z)$.

In the geometry shown in fig.~1, the optical operator for the
exciton in the dot 1 can be written as follows:
$\hat{V}^{opt}_{1,+}={\bf e}_1\hat{{\bf
p}}=\cos(\theta_1)\hat{p}_x+i\hat{p}_y+\sin(\theta_1)\hat{p}_z$,
where $\hat{{\bf p}}$ and ${\bf e}_1$ are the momentum operator
and polarization vector, respectively. Using this operator, the
probability of inter-band optical transitions for the dot 1 takes
a form
$P^{s,\mu}_1=|<\Psi_1^{s,n,l}|\hat{V}_{1,+}^{opt}|\Psi_1^{\mu,n,l}>|^2$,
where $s=\uparrow(\downarrow)$ and $\mu=\pm3/2,\pm1/2$. Emission
of the dot 2 is described in a similar way with the operator
$\hat{V}^{opt}_{2_\pm}=\cos(\theta_2)\hat{p}_x\pm
i\hat{p}_y-\sin(\theta_2)\hat{p}_z$, where signs $\pm$ relate to
the different polarizations of the secondary photon ($\mp\hbar$).
For simplicity, we consider the case when the linear momenta of
both photons lie in the $x-z$ plane.

The inter-dot transfer is described by the Coulomb operator which
can be expanded into an infinite series of multipole terms.
However, it is natural to assume that the dipole-dipole
interaction will provide the leading term,

\begin{eqnarray}
\label{Coul} \hat{V}_{Coul}=\frac{e^2}{\epsilon R^3}({\bf r}_1{\bf
r}_2-3z_1z_2),
\end{eqnarray}
where ${\bf r}_{1(2)}$ are the radius vectors related to the dots
(fig.~1a), $\epsilon$ is the averaged dielectric constant, and $R$
is the distance between the dots. Below, we will generalize our
results including multipole interactions. The F\"{o}rster-like
probability of inter-dot transition takes the form

\begin{eqnarray}
\label{Forster} W_{\beta_1}= \frac{2\pi}{\hbar}\sum_{\beta_2}
|<\beta_1|\hat{V}_{Coul}|\beta_2>|^2
\delta(E_{\beta_1}-E_{\beta_2}),
\end{eqnarray}
where the indices $\beta_{1(2)}$ denote the exciton states in the
dots: $\beta_1=(s_1,\mu_1,n_1,l_1)$ and
$\beta_2=(\mu_2,s_2,n_2,l_2)$. Because of fast intra-dot energy
relaxation, the function $|\beta_1>$ in eq.~\ref{Forster}
describes the ground-state exciton in the dot 1 with
$s_1=\uparrow,\downarrow$ and $\mu_1=\pm3/2$, and
$(n_1,l_1)=(1,0)$. In the spirit of the F\"{o}rster theory the
delta function in eq.~\ref{Forster} should be replaced by the
spectral overlap integral $J_{\beta_1,\beta_2}=\int
\rho_{\beta_1}(E)\rho_{\beta_2}(E)dE$ which involves normalized
line shapes
$\rho_{\beta_i}(E)=\pi^{-1}\Gamma_{\beta_i}/[(E-E_{\beta_i})^2+\Gamma_{\beta_i}^2]$,
where $\Gamma_{\beta_i}$ is the homogeneous broadening of the
exciton $\beta_i$. Lorentzians were utilized for simplicity.

By using eqs.~\ref{WF1}-\ref{Forster}, we now compute the mean
spin in the dots and the degree of polarization of secondary
photons. To be specific, we consider the resonant dipole-allowed
absorption process of incident photon in the dot 1 that involves a
heavy-hole level (fig.~1b); in other words, the incident-photon
energy is taken below the first inter-band transition related to
the light hole. The mean $z$-component of electron spin
polarization in the dot 1 is determined by the probabilities
$P^{s,\mu}_1$ and is equal to
$S_1=(P_1^\uparrow-P_1^\downarrow)/(P_1^\uparrow+P_1^\downarrow)=
-2\cos(\theta_1)/[\cos(\theta_1)^2+1]$, where
$P_i^{\uparrow(\downarrow)}\propto|P_{cv}|^2(cos(\theta_1)\mp1)^2$
is the probability of the electron being in the state
$\uparrow(\downarrow)$ and $P_{cv}=<S|\hat{p}_x|X>$ is the
inter-band optical matrix element. In the optical matrix elements,
the operator ${\bf \hat{p}}$ was involved only in the integrals
with the Bloch functions. For the next step, we calculate the
Coulomb matrix elements under resonance conditions. In the regime
of inter-dot resonance, the ground-state exciton energy of the dot
1 is equal to the energy of excited dipole-active exciton in the
dot 2. The latter state can be composed of either heavy-hole or
light-hole. We start with the resonance between heavy-hole states
in the dots (fig.~1b). The probability to create the exciton with
$s_2=\uparrow$ in the dot 2 is given by
$P^{\uparrow}_2=(1/2)P_1^\uparrow
W_{\beta_1\rightarrow\beta_2}=P_1^\uparrow
w_0J_{\beta_1,\beta_2}/2$, where $\beta_1=(\uparrow,3/2,1,0)$ and
$\beta_2=(\uparrow,3/2,n_2,l_2)$; the factor $1/2$ is the
probability to find the heavy hole in either state ($\pm3/2$) in
the dot 1; this is due to fast momentum relaxation of holes.
Besides, $W_{\beta_1\rightarrow\beta_2}$ is the probability of
F\"{o}rster-like transfer between the states $\beta_1$ and
$\beta_2$. A coefficient $w_0=2\pi d_0^4(e^4/\epsilon^2R^6\hbar)$,
where $d_0=<X|x|S>$ is the atomic dipole moment. For the spin
$\downarrow$ we have a similar equation,
$P^{\downarrow}_2=P_1^\downarrow w_0J_{\beta_1,\beta_2}/2$. Again,
the operator ${\bf r}$ was involved only in the integrals with the
Bloch functions. The spin polarization of the dot II is given by

\begin{eqnarray}
\label{SII'}
S_2'=\frac{P_2^\uparrow-P_2^\downarrow}{P_2^\uparrow+P_2^\downarrow}=S_1=
-\frac{2\cos(\theta_1)}{[\cos(\theta_1)^2+1]}.
\end{eqnarray}
If $\theta_1=0$, the system has axial symmetry, the transfer
process conserves the total momentum, and therefore $S_2'=-1$.
Thus, F\"{o}rster transport {\it preserves the spin polarization}
in the regime of inter-dot resonance between heavy-hole levels.
Now we assume that the parameters of dots are chosen to satisfy
the condition of inter-dot resonance between heavy and light holes
(fig.~1c). It is easy to see that Coulomb transfer results in the
{\it spin flipping}. For example the probability $P^{\uparrow}_2$
is now expressed via $P^{\downarrow}_1$: \
$P^{\uparrow}_2=(1/2)P_1^\downarrow
W_{\beta_1\rightarrow\beta_2}=P_1^\downarrow
w_0J_{\beta_1,\beta_2}/6$, where $\beta_2=(\uparrow,1/2,n_2,l_2)$.
Similarly, $P^{\downarrow}_2=P_1^\uparrow
w_0J_{\beta_1,\beta_2}/6$. Thus, we obtain the effect of spin
flipping:

\begin{eqnarray}
\label{SII''} S_2''=-S_1.
\end{eqnarray}
So far, we considered strongly resonant conditions. In the general
case, the mean spin in the dot 2 is calculated as

\begin{eqnarray}
\label{Stot} S_2=\frac{\sum_{\beta_1,\beta_2}
P^{\uparrow}_2(\beta_1\rightarrow\beta_2)-P^{\downarrow}_2(\beta_1\rightarrow\beta_2)}
{\sum_{\beta_1,\beta_2}P^{\uparrow}_2(\beta_1\rightarrow\beta_2)+P^{\downarrow}_2(\beta_1\rightarrow\beta_2)},
\end{eqnarray}
where the summation involves all pairs of states; the index
$\beta_1$ is related to the $hh$ ground state of the dot 1:
$\beta_1=(\uparrow(\downarrow),\pm3/2,1,0)$. The degree of
circular polarization of secondary photons at the dot-2
ground-exciton energy is now written as

\begin{eqnarray}
\label{Polar} P_{circ}=\frac{I_+-I_-}{I_++I_-}=
-S_2\frac{2\cos(\theta_2)}{[\cos(\theta_2)^2+1]},
\end{eqnarray}
where $I_{\pm}$ are the light intensities given by
$I_{+}=P_2^{\uparrow}P_+(\uparrow)+P_2^{\downarrow}P_+(\downarrow)$
and
$I_{-}=P_2^{\uparrow}P_-(\uparrow)+P_2^{\downarrow}P_-(\downarrow)$.
Here, the optical transition rate $P_{\sigma}(s)$ describes the
emission process in which an electron with the spin $s$ in the dot
2 creates a photon with the circular polarization $\sigma$, where
$\sigma$ can be $+$ or $-$. The degree of circular polarization
(\ref{Polar}) strongly depends on the resonance conditions between
the QDs (fig.~2a). If $\theta_{1(2)}=0$, the system has axial
symmetry and the electron spin is either conserved or flipped in
the resonant-transfer process (fig.~2a). The latter comes from the
conservation of the total angular momentum in the Coulomb matrix
elements. Besides,  the rate of exciton transfer,
$1/\tau_{trans}=W_{\beta_1}$, is strongly enhanced under the
inter-dot resonance conditions (fig.~2b). Note that the total
angular momentum is not conserved in the three-step process shown
in figs.~1b,c because of fast relaxation of angular momentum for
the hole.

{\it Spherical quantum dots with a cubic lattice. } In sherical
dots, symmetry of a single QD is high and both heavy and light
holes will contribute to the transfer rate for the given inter-dot
resonance. The multi-component wave functions for the holes in a
model with infinite walls are well known \cite{Efros-PR96},
\begin{eqnarray}
\label{WF-sph} \Psi_i^M=\sum_{l,m,\mu}
C_{l,m,\mu,M}R^{(i)}_l(r)Y^{(i)}_{lm}(\Omega)u_{\mu}.
\end{eqnarray}
Here $i$ is the QD number ($i=1,2$), $M$ is the $z$-component of
total angular momentum, $Y^{(i)}_{lm}(\Omega)$ are spherical
harmonic functions, $R^{(i)}_l(r)$ are functions of radial motion
\cite{Efros-PR96}, and $\mu=\pm1/2,\pm3/2$. Calculation of the
spin orientation in the dots 1 and 2 is straightforward. The mean
$z$-components of spin in dots are written as
$S_1=-\cos(\theta_1)/2$ and $S_2=S_1/2$. The degree of circular
polarization of emitted light takes a form $P_{circ}=-S_2
\cos(\theta_2)/2=\cos(\theta_1)\cos(\theta_2)/8$. At the angles
$\theta_{1(2)}=0$, the polarization of emitted light is maximal
and equal to $P_{circ}=1/8$. The degree of polarization, $1/8$,
appears as a result of the three-step process. According to the
theory of spin orientation in 3D crystals, the degree of
polarization in the two-step process is $1/4$
\cite{optical-orientaion}. Since the band structure of cubic
spherical dots is isotropic, electron spin transfer does not
depend on the type of inter-dot resonance and the electron spin is
not flipped.

{\it Oblate quantum dots with cubic and wurtzite lattices.}
Quantum dots can be anisotropic due to both shape and crystal
lattice. Such anisotropy strongly affects the valence band
structure giving rise to splitting between heavy- and light-hole
levels. In nearly spherical crystals, anisotropy can be taken into
account with perturbation theory \cite{Efros-PR96}. The 4-fold
degeneracy of the hole states is split into two 2-fold degenerate
states. The splitting can be written as
$\Delta=\Delta_{cr}+\Delta_{shape}$, where $\Delta_{cr}$ is the
crystal field splitting in a hexagonal lattice (like in CdSe) and
$\Delta_{shape}$ is the splitting due to the shape. The Kramers
doublet of hole states has the quantum numbers $|M|=1/2$ and
$|M|=3/2$. First, we consider two oblate dots forming a molecule
with axial symmetry. In such a molecule, ${\bf c}_1||{\bf
c}_2||{\bf z}$, where ${\bf c}_{1(2)}$ are the symmetry axes of
dots (fig.~3~b). To be more specific, we assume that the ground
state of holes have the angular momenta $|M|=3/2$, like in the
dots based on InP. Using the wave functions (\ref{WF-sph}) it is
easy to show that all results for disk-shaped QDs hold in the case
of oblate dots with ${\bf c}_1||{\bf c}_2||{\bf z}$.

{\it Randomly-oriented QD pairs.} \ It is natural to suppose that
the randomness of nanocrystal axes in a QD solid will change the
spin transfer rates. To calculate the spin transport rates in a
pair of arbitrary-oriented dots, one can use the matrices of
rotation for spin and spatial functions \cite{Landau} and
introduce Eulerian angles for the dots,
$\phi_i^{(1)},\phi_i^{(2)},\phi_i^{(3)}$, where $i=1,2$ (fig.~3b).
By using the matrices of rotation, the coordinate system $(x,y,z)$
is transformed into the systems $(x'_i,y'_i,z'_i)$ where
individual dots have symmetry of oblate ellipsoids.  The spin
transfer probabilities for oblate dots depend only on the angles
$\phi_i^{(1)}$ and $\phi_i^{(2)}$. Under resonance conditions the
mean spins in the dots are connected by equation

\begin{eqnarray}
\label{angle} S_2=S_1\frac{W_a-W_b}{W_a+W_b},
\end{eqnarray}
where $W_a=
W_{\uparrow\rightarrow\uparrow}=W_{\downarrow\rightarrow\downarrow}$
and
$W_b=W_{\uparrow\rightarrow\downarrow}=W_{\uparrow\rightarrow\downarrow}$.
The coefficients $W_{a(b)}$ describe probabilities of inter-dot
transitions with conservation (flipping) of spin and are
complicated functions of $\phi_i^{(1)}$ and $\phi_i^{(2)}$. In a
system with randomly-oriented molecules, spin transfer does not
vanish; it can seen by calculating averaged probabilities
$\bar{W}=<W>_{\phi_1^{(1)},\phi_1^{(2)},\phi_2^{(1)},\phi_2^{(2)}}$
and spins
$\bar{S}_i=<S_i>_{\phi_1^{(1)},\phi_1^{(2)},\phi_2^{(1)},\phi_2^{(2)}}$.
The ratio between averaged probabilities $\bar{W}_a/\bar{W}_{b}$
depend on the type of inter-dot resonance:
$\bar{W}_a/\bar{W}_{b}=1.61$ for the $hh-hh$ resonance and
$\bar{W}_a/\bar{W}_{b}=1/1.61$ for the resonance between $hh$ and
$lh$ states. Thus, the $hh-hh$ transfer conserves partially the
spin orientation, whereas the $hh-lh$ inter-dot coupling leads to
the flipping of spin. For the case shown in fig.~3b,
$\theta_1=\theta_2=0$ and the calculated mean spin in the dot~1 is
given by $\bar{S}_1=-0.586$. The dot-2 spin becomes
$\bar{S}_2=-0.22$ and $0.22$ in the case of $hh-hh$ and $hh-lh$
resonances, respectively. Experimentally, the spin orientation in
the dot~2 can be observed by measuring the degree of circular
polarization of secondary photons. We find that
$\bar{P}_{circ}^{hh-hh}=0.13$ and $\bar{P}_{circ}^{hh-lh}=-0.25$.

{\it Quantum-dot chains.} If cylindrical dots form an
ideally-oriented chain (like in self-assembled monolayers,
fig.~3a) and all of the dots are under resonance conditions, the
spin can be transferred along the chain without losses, $S_N=\pm
S_1$, where $S_1$ and $S_N$ are the mean spins in the first and
$N^{th}$ dots, respectively (fig.~3c). The sing $\pm$ in the above
relation depends on the types of inter-dot resonances. If an ideal
chain is formed of spherical dots, the transferred spin rapidly
decreases with the number of dots, $S_N=S_1/2^N$. In disordered
chains, there is an additional mechanism of spin randomization.
For oblate crystals with randomly-oriented axes and under
inter-dot resonance conditions, we can estimate the decay of spin
using averaged probabilities, $\bar{W}_{a,b}$. This leads to
$S_N\sim0.2^N S_1$.

\vskip0.5 cm

The dipole-dipole interaction (\ref{Coul}) provides the main
contribution to the transfer rate. At the same time, the higher
multipole terms of the Coulomb operator can certainly affect the
magnitude of transfer rate and lead to additional inter-dot
resonances which should be consistent with symmetry. However, the
spin-transfer selection rules established above will hold beyond
the dipole-dipole approximation because these rules come from
axial symmetry in a QD pair. Specifically, the $hh-hh$ and $hh-lh$
inter-dot resonances will result in conservation and flipping of
spin, respectively.

Experimentally, the most preferable systems to observe spin
transport are the system with QD monolayres \cite{exper2,exper3}
or a single QD pair \cite{exper3} on a surface. In the first case,
all QD pairs have the same orientation of the molecular axis ${\bf
R}$ (fig.~3a). If QDs are spherical, the spin polarization in the
dots 2 will be $S_1/2$.  In the case of oblate QDs with
randomly-oriented QD axes, the spin orientation will remain
non-zero under inter-dot resonance conditions. Another suitable
system is a single QD molecule bound to a surface which can be
studied by available methods of single-dot spectroscopy
\cite{Bawendi}.

Another important issue related to exciton transport is the
strength of dipole transitions. In this paper, we assumed that the
ground-state excitons in dots are optically active. This would not
be the case for CdSe dots where a strong inter-band exchange
interaction splits exciton levels. The resulting exciton ground
state turns to be dark. Our results are fully applicable to the
QDs with optically-active excitons. For example, excitons in the
ground state are optically-active in InP nanocrystals where the
exchange interaction is weak \cite{InP}. In addition, the exciton
ground states are optically active in self-assembled QDs
\cite{stachedQDs}. Such self-assembled QDs are usually
lens-shaped. The case with optically-inactive ground states of
excitons should be considered specially. In addition, we
considered excitons within the single-particle approximation
ignoring the intra-dot Coulomb interaction. This approximation is
justified for our dot parameters since the typical energy of
in-plane quantization is greater than the intra-dot Coulomb
interaction \cite{Govorov}.

A moderate magnetic field can favor the observation of spin
transport because it induces the spin-splitting and strongly
enlarges the degree of circular polarization of emitted light in
QDs \cite{B-spins,B-QDs}. In the system with monolayers, the
magnetic field can be applied parallel to the molecular axis ${\bf
R}$. To observe spin transfer between QDs, one should have a
sufficiently long spin-relaxation time. The spin-relaxation times
found in experiments on the bulk semiconductors and QDs range from
$100~ps$ to $100~\mu s$ \cite{Awschalom1,B-spinsQD1,B-spinsQD2}.
The exciton-transfer times in nanocrystals, recently measured in
refs.~\cite{exper1,exper2}, are in the range from $700~ps$ to
$10~ns$. This tells us that suitable conditions to observe spin
transport of electrons can be found experimentally. By analyzing
the rate equations, one can see that the mean spin in the dot 2
depends mostly on the ratio $\tau_{trans}/\tau_{e-spin}$, where
$\tau_{e-spin}$ is the spin relaxation time for the electron and
$\tau_{trans}$ is the inter-dot transfer time of excitons. At the
same time, the emission intensity of the dot-2 is determined by
the ratio $\tau_{trans}/\tau_{exc}$. Spin and energy transfers
become efficient if $\tau_{trans}\leq \tau_{e-spin}$ and
$\tau_{trans}\leq \tau_{exc}$, respectively. The latter was
satisfied in recent experiments \cite{exper1,exper2,exper3}.

The rate of energy transfer between QDs can strongly depend on
temperature and resonance conditions. We now assume that the QD
pair is designed to satisfy the resonant condition,
$E_1^0=E_2^{exc}$, where $E_1^0$ is the ground-state energy of
exciton in  the dot 1 and $E_2^{exc}$ is related to the excited
exciton state in the dot 2. In the case of $hh-hh$ inter-dot
resonance, the transfer time can be estimated as
$\tau_{trans}=1/(w_0J)$. Here we will use the parameters of InP:
$d_0=6~\AA$ and $\epsilon=12.6$. At low temperatures, homogeneous
broadenings of excitons are relatively small and
$\Gamma_1^0\ll\Gamma_2^{exc}$, where $\Gamma_1^0$ and
$\Gamma_2^{exc}$ are the broadenings of exciton levels in the dots
1 and 2, respectively. We obtain $\tau_{trans}\sim 120~ps$ taking
parameters $\Gamma_1^0=1~meV$, $\Gamma_2^{exc}=5~meV$, and
$R=70~\AA$. At room temperature, we find $\tau_{trans}\sim 1~ns$
with $\Gamma_1^0\sim\Gamma_2^{exc}\sim20~meV$.

To calculate the spin orientation in nanosrystals, we assumed that
the time of momentum relaxation for the holes is much shorter that
the spin-relaxation time for the electrons. This relation is
typical for experiments. The momentum relaxation time of holes in
solids and nanostructures is often short due to strong $hh-lh$
mixing in the valence band and relatively weak quantization of
energy levels of holes \cite{Ivchenko}.

To conclude, we have studied spin transfer in nanocrystals which
does not involve transport of charge. It has been demonstrated
that the spins can be efficiently transferred between quantum dots
via the Coulomb interaction. In the transfer process the electron
spin can be conserved or flipped. The transferred spin
polarization survives even in randomly-oriented QD pairs and
chains.

The author acknowledges Garnett Bryant for helpful discussions on
optical properties of quantum dots. This work was supported by the
Condensed Matter and Surface Science Program at Ohio University
and by the Volkswagen Foundation.

\begin{figure}
\includegraphics[width=4.00in,angle=0]{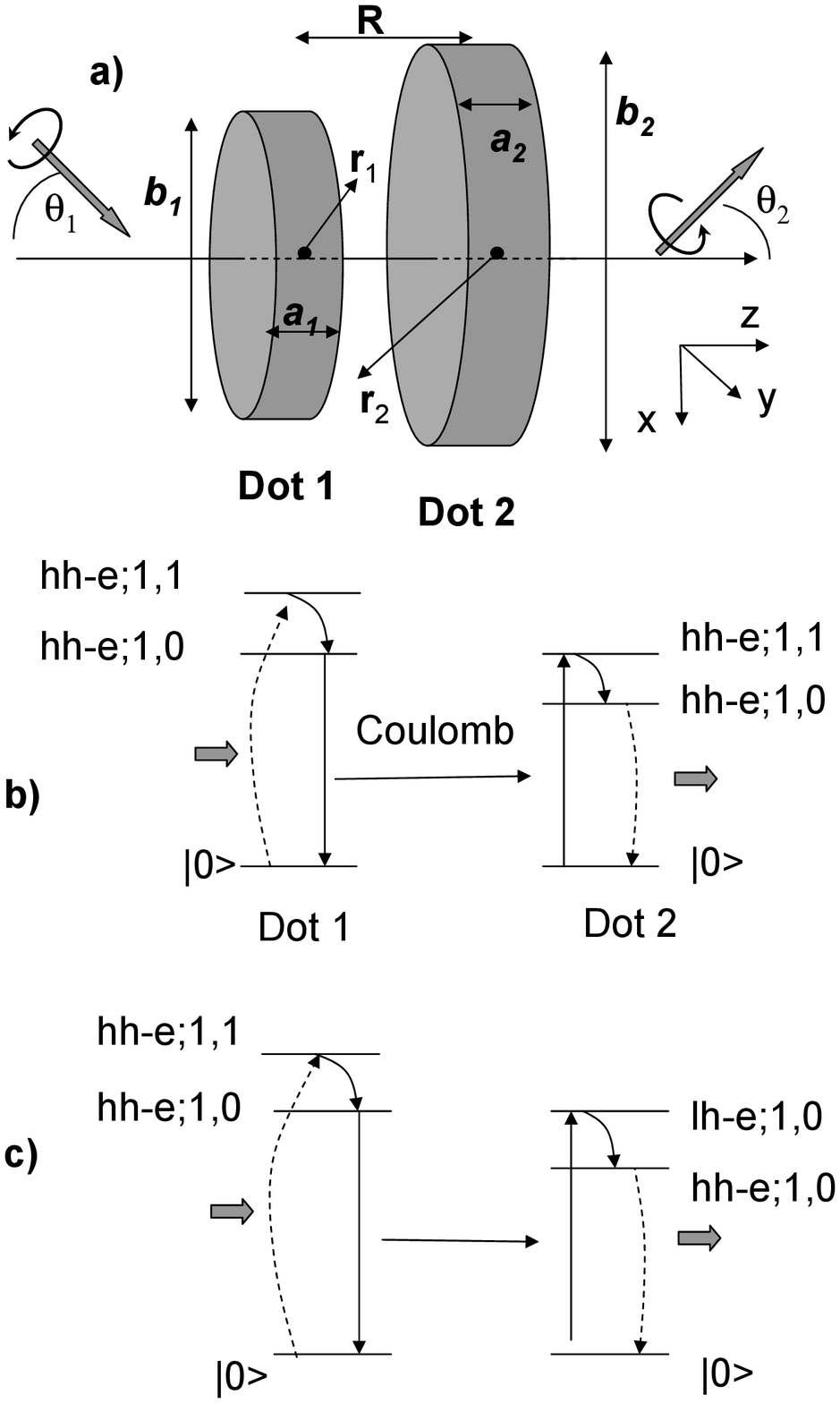}
\caption{\label{fig1}  \ \\ \ Sketch of the quantum-dot molecule
(a). Energy diagrams of intra- and inter-dot transitions in the
transfer process (b and c); the label $(\gamma-e;n,l)$ denotes the
exciton composed of hole $\gamma$ and electron, where $\gamma$ can
be $hh$ or $lh$, and $(n,l)$ are the envelope-function indices.}
\end{figure}

\begin{figure}
\includegraphics[width=5.00in,angle=-90]{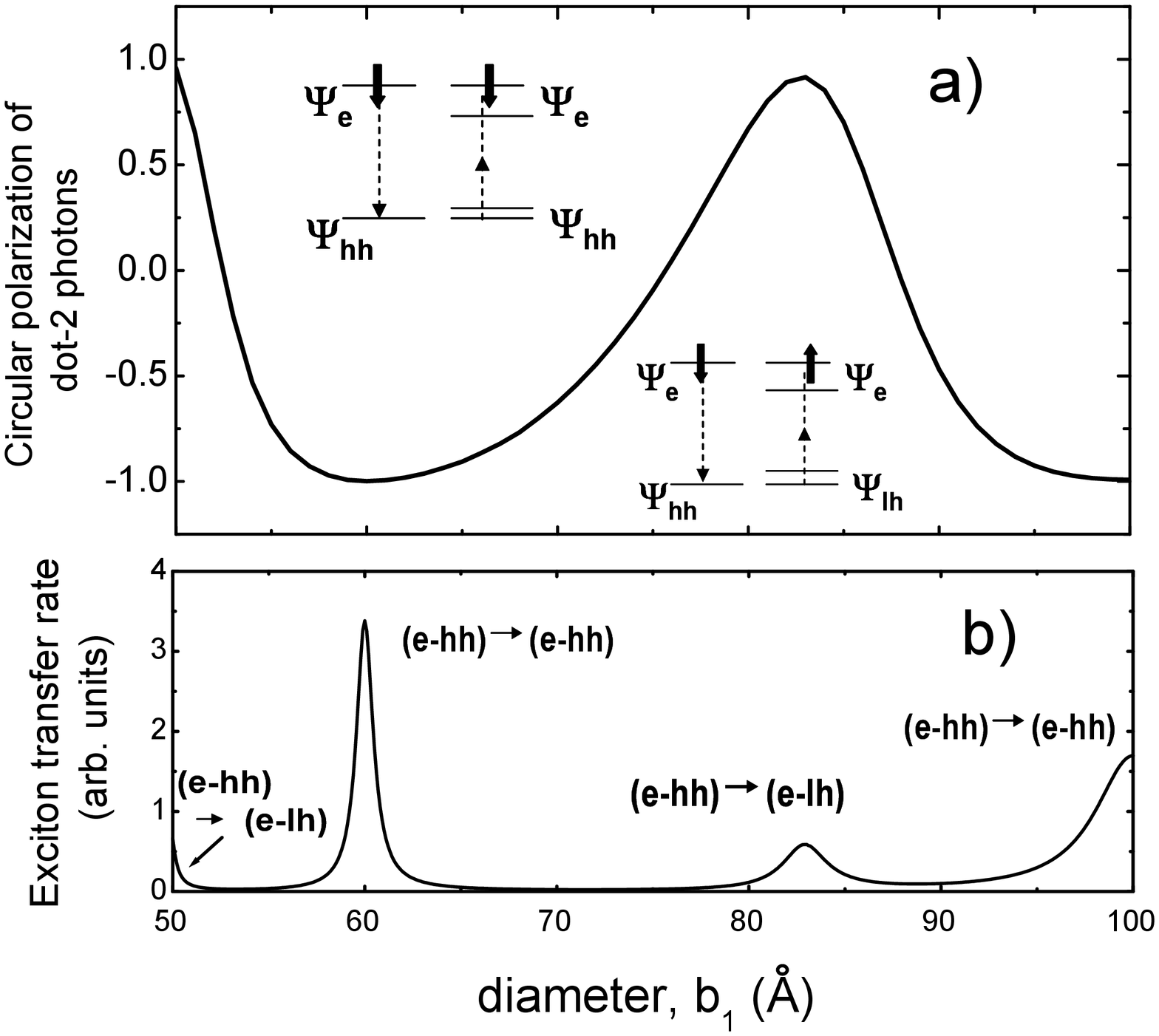}
\caption{\label{fig1}  \ \\ \ (a) Calculated degree of circular
polarization of photons emitted by the dot 2 as a function of the
dot-1 diameter; $\theta_{1(2)}=0$. The sizes of the dot 2 are kept
constant, whereas the diameter of the dot 1 is varied. (b)
Calculated rate of exciton transfer from the dot 1 into dot 2.
Inserts show diagrams of inter-band transitions. The crystal
parameters correspond to InP quantum dots; effective masses:
$m_e=0.077m_0$, $m_{lh}=0.12m_0$, $m_{hh}=0.6m_0$; $R=80~\AA$,
$a_1=a_2=25~\AA$, $b_2=100~\AA$, and $50<b_1<100~\AA$. The
low-temperature broadening of the ground state of exciton in the
dot 1 is taken as $1~meV$; the broadening of all excited states in
the dot 2 is assumed to be $5~meV$. }
\end{figure}

\begin{figure}
\includegraphics[width=5.00in,angle=0]{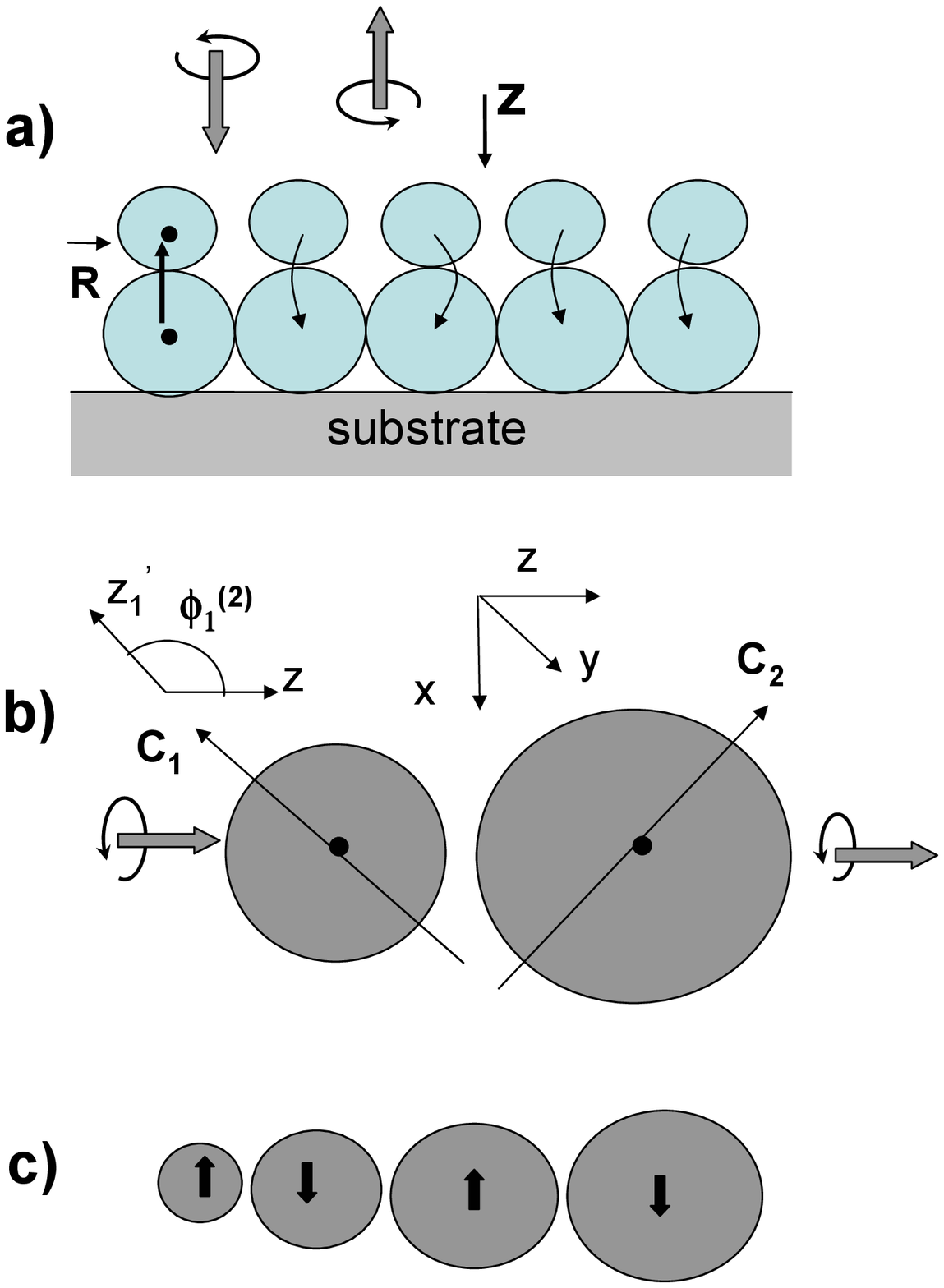}
\caption{\label{fig1} \ \\ \ (a) Schematic of the system with two
monolayers of dots. Similar systems were studied experimentally in
refs.~9,10. (b and c) Sketches of a pair of randomly-oriented dots
and a quantum-dot chain. }
\end{figure}


\begin{thebibliography}{10}

\expandafter\ifx\csname bibnamefont\endcsname\relax
     \def\bibnamefont#1{#1}\fi
\expandafter\ifx\csname bibfnamefont\endcsname\relax
     \def\bibfnamefont#1{#1}\fi
\expandafter\ifx\csname url\endcsname\relax
     \def\url#1{\texttt{#1}}\fi
\expandafter\ifx\csname
urlprefix\endcsname\relax\def\urlprefix{URL }\fi
\expandafter\ifx\csname bibinfo\endcsname\relax
\def\bibinfo#1#2{#2}\fi \expandafter\ifx\csname
eprint\endcsname\relax \def\eprint#1{#1}\fi


\bibitem{spins1}
G. A. Prinz, Science 282, 1660 (1998);  Y. Ohno, D. K. Young, B.
Beschoten, F. Matsukura, H. Ohno, D.D.Awschalom, Nature 402, 790
(1999); R. M. Potok, J. A. Folk, C. M. Marcus, and V.Umansky,
Phys. Rev. Lett. 89, 266602 (2002).

\bibitem{spins2}
S. Datta and B. Das, Appl. Phys. Lett. 56, 665 (1990); E. I.
Rashba, Phys. Rev. B 62, R16267  (2000) ; G. Schmidt, D. Ferrand,
L. W. Molenkamp, A. T. Filip, and B. J. van Wees, Phys. Rev. B 62,
R4790 (2000).

\bibitem{Ganichev} S. D. Ganichev, E. L. Ivchenko,
V. V. Bel'kov, S. A. Tarasenko, M. Sollinger, D. Weiss, W.
Wegscheider, and W. Prettl, Nature {\bf 417}, 153 (2002).

\bibitem{QComp}
D. Loss and D. P. DiVincenzo, Phys. Rev. A 57, 120 (1998).

\bibitem{optical-orientaion} F. Meier and B.P. Zakharchenya, (eds),
{\it Optical Orientation} (North-Holland, Amsterdam, 1984).

\bibitem{Dyakonov} M. I. D'yakonov and V. I. Perel',
 Sov. Phys. JETP 33, 1053 (1971)[Zh. Exper. Teor. Fiz. 60, 1954
 (1971)].

\bibitem{Awschalom1} J. M. Kikkawa and D. D. Awschalom, Phys. Rev.
Lett. 80, 4313 (1998).

\bibitem{exper1}
C. R. Kagan, C. B. Murray, and M. G. Bawendi, Phys. Rev. B 54,
8633 (1996); A. Javier, C. S. Yun, J. Sorena, and G. F. Strouse,
J. Phys. Chem. B, 107, 435 (2003).

\bibitem{exper2} S. A. Crooker, J. A. Hollingsworth, S. Tretiak, and V. I.
Klimov, Phys. Rev. Lett. 89, 186802 (2002).

\bibitem{exper3}
A.A. Mamedov, A. Belov, M. Giersig, N.~N.~Mamedova, and
N.~A.~Kotov, J. Am. Chem. Soc. 123, 7738 (2001); S. Weng, N.
Mamedova, N. A. Kotov,  W. Chen,  and J. Studer, NanoLetters 2,
817 (2002).

\bibitem{forster}
Th. F\"{o}rster, in {\it Modern Quantum Chemistry}, edited by O.
Sinanoglu (Academic, New York, 1965), pp. 93-137; D.~L.~Andrews
and A.~A.~Demidov (eds.), {\it Resonance Energy Transfer  }
(Wiley, N.Y., 1999).

\bibitem{theory} G. W. Bryant, Physica B 314, 15 (2002);
B.~Lovett, J. H. Reina, A. Nazir, B. Kothari, and A. Briggs,
quant-ph/0209078.

\bibitem{stachedQDs}
A. A. Darhuber, V. Holy, J. Stangl, G. Bauer, A. Krost, F.
Heinrichsdorff, M. Grundmann, D. Bimberg, V. M. Ustinov and P. S.
Kop'ev, A. O. Kosogov,  and P. Werner, Appl. Phys. Lett. 70, 955
(1997); I. Shtrichman, C. Metzner, B. D. Gerardot, W. V.
Schoenfeld, and P. M. Petroff, Phys. Rev. B 65, 081303 (2002).

\bibitem{Bawendi}
K. T. Shimizu, W. K. Woo, B. R. Fisher, H. J. Eisler, and M. G.
Bawendi, Phys. Rev. Lett. 89, 117401 (2002).

\bibitem{Ivchenko}
E. L. Ivchenko and G. E. Pikus, {\it Superlattices and Other
Heterostructures. Symmetry and Optical Phenomena} (Springer,
Berlin, 1997).


\bibitem{Efros-PR96} Al. L. Efros, M. Rosen, M. Kuno,
M. Nirmal, D. J. Norris, and M. Bawendi, Phys. Rev. B 54, 4843
(1996)

\bibitem{Landau} L. D.Landau, and E. M. Lifshitz,
{\it Quantum Mechanics: Non-Relativistic Theory, Vol.3} (Pergamon
Press, Oxford, 1981).

\bibitem{B-spins}
W. Heller and U. Bockelmann, Phys. Rev. B 55, R4871 (1997); M.
Bayer, A. Kuther, A. Forchel, A. Gorbunov, V. B. Timofeev, F.
Schäfer, J. P. Reithmaier, T.L.Reinecke, and S.N.Walck, Phys. Rev.
Lett. 82, 1748 (1999); D. Gammon and D. G. Steel, Phys. Today 55,
36 (2002); M. Paillard, X. Marie, P. Renucci, T. Amand, A. Jbeli,
and J. M. Gerard, Phys. Rev. Lett. 86, 1634 (2001).

\bibitem{B-QDs}
C. Schulhauser, D. Haft, R. J. Warburton, K. Karrai, A.O. Govorov,
A. V. Kalameitsev,  A. Chaplik, W.Schoenfeld, J. M. Garcia, and P.
M. Petroff, Phys. Rev. B 66, 193303 (2002).

\bibitem{InP} R. J. Ellingson, J. L. Blackburn, J. Nedeljkovic,
G. Rumbles, M. Jones, H. Fu, and A. J. Nozik, Phys. Rev. B 67,
075308 (2003).

\bibitem{Govorov}
A.~O.~Govorov and A.~V.~Chaplik, Sov. Phys. JETP, 72, 1037 (1991).

\bibitem{B-spinsQD1}  J. A. Gupta, D. D. Awschalom, Al. L. Efros, and A. V. Rodina,
Phys. Rev. B 66, 125307 (2002).

\bibitem{B-spinsQD2} V. D. Kulakovskii, K. Babocsi, M. Schmitt,
N. A. Gippius, and W. Kiefer, Phys. Rev. B 67, 113303 (2003).



\end{thebibliography}
\end{document}